\documentclass[12pt]{iopart}

\usepackage{iopams}  
\usepackage[dvips]{graphics}
\newcommand{\pc}{\mathbb{P}}
\newcommand{\reals}{\mathbb{R}}

\newcommand{\urries}[1]{[\hspace{-1pt}[ #1 ]\hspace{-1pt}]}
\newcommand{\julve}[2]{\langle #1 #2 \rangle}

\begin{document}

\title{Pauli-Villars regularisation and Born-Infeld kinematics}

\author{Frederic P Schuller\dag, Mattias N R Wohlfarth\dag\ \\ and Thomas
  W Grimm\ddag}

\address{\dag\ Department of Applied Mathematics and Theoretical
  Physics, \\Centre for Mathematical Sciences, University of Cambridge,
  \\ Wilberforce Road, Cambridge CB3 0WA, UK}

\address{\ddag\ II. Institut f\"ur Theoretische Physik, Universit\"at
   Hamburg, \\ 22761 Hamburg, Germany}

\begin{abstract}
  Dynamical symmetries of Born-Infeld theory can be absorbed into the
  spacetime geometry,
  giving rise to relativistic kinematics with an additional invariant
  acceleration scale.
  The standard Poincar\'e group $\mathcal{P}$ is thereby enhanced to its
  pseudo-complexified version, which is isomorphic to
  $\mathcal{P}\times \mathcal{P}$. We construct the irreducible representations
  of this group, 
  which yields the particle spectrum of a relativistic quantum theory
  that respects a maximal acceleration. It is found that each
  standard relativistic particle is associated with a
  'pseudo'-partner of equal spin but generically different mass.
  These pseudo-partners act as Pauli-Villars
  regulators for the other member of the doublet, as is found from the explicit
  construction of quantum field theory on pseudo-complex spacetime.
  Conversely, a Pauli-Villars regularised quantum field theory on real
  spacetime possesses a field phase space with integrable
  pseudo-complex structure, which gives rise to a quantum field theory
  on pseudo-complex spacetime.

  This equivalence between  maximal acceleration
  kinematics,  pseudo-complex quantum field theory, and
   Pauli-Villars regularisation rigorously establishes a conjecture on the regularising
  property of the maximal acceleration principle in quantum field
  theory, by Nesterenko, Feoli, Lambiase and Scarpetta.\\[24pt]
  \textsl{Journal Reference:} Classical and Quantum Gravity (to appear)
\end{abstract}

\pacs{03.70.+k, 03.65.Pm, 11.30.Cp}


\maketitle

\section{Introduction}
Standard quantum field theory is built on the assumption that
fundamental particles are irreducible representations of the spacetime
Poincar\'e group. The physical rationale for this is the
assumption that the symmetries of Maxwell electrodynamics, associated
with an invariant speed $c$, are
kinematical, i.e., all fundamental theories must possess them.
However, quantum field theory is plagued with divergences, as is
Maxwell electrodynamics of point particles. Indeed, the problem of
infinite electric field energy of a charged point particle led to
the formulation of Born-Infeld electrodynamics \cite{Born:1934gh}. The latter parameterises
the energy divergence, but preserves Lorentz invariance. This
raises the question of whether replacing the kinematics of Maxwell
theory, i.e., special relativity, by extended kinematics extracted from
Born-Infeld theory, might also regulate the divergences encountered
in quantum field theory. In this paper we show that this indeed is
the case.

This result is of particular interest, since Dirac-Born-Infeld theory
presents the electrodynamics on a Dirichlet brane \cite{Abu87} in
ten-dimensional 
superstring theories. In this context, the maximal acceleration is
recognised as the inverse fundamental string length. Brane world
scenarios, e.g. \cite{Randall}, assign to Dirichlet three-branes the r\^ole
of the observed four-dimensional universe. This implies
kinematical consequences for physical observers  in such
models, as it was shown in \cite{Schuller:2002rr,Schuller:2002fn} that the kinematisation of the
symmetries associated with the maximal field strength $b^{-1}$
 of Born-Infeld theory,
\begin{equation}\label{BIoriginal}
  L_{BI} = {\det}^\frac{1}{2}\left(\eta_{\mu\nu} + b
  F_{\mu\nu}\right),
\end{equation}
leads to the pseudo-complexified Lorentz group $SO_\pc(1,3)$ on
pseudo-complexified Minkowski spacetime $\pc^{1,3}$, where the
commutative ring of
pseudo-complex numbers is defined as
\begin{equation}
  \pc \equiv \{a+Ib | a,b \in \reals, I^2=+1, I\not\in \reals\},
\end{equation}
as will briefly be reviewed in section \ref{sec_pcM}.
Resulting corrections to symmetry-sensitive calculations in
relativity, such as the Thomas
precession \cite{Thomas}, yield an upper bound
\cite{Schuller:2002rr} for the Born-Infeld parameter,
\begin{equation}
  b \leq 10^{-11} C N^{-1},
\end{equation}
or, equivalently, a lower bound
\begin{equation}\label{maxaccel}
  \mathfrak{a} \geq 10^{22} m s^{-2}
\end{equation}
on the maximal acceleration
of an electrically charged particle coupled to Born-Infeld theory, in
order to be in accordance with data from high precision
experiments \cite{Newman}. 
The real Lorentz group is a subgroup of $SO_\pc(1,3)$. Therefore, 
Born-Infeld kinematics, encoded in the pseudo-complex Lorentz group,
extends special relativity \textsl{without} deforming
Lorentz invariance, in contrast to other approaches to theories with
two fundamental constants \cite{Amelino-Camelia,Kowalski-Glikman:2002ft}. 
Most importantly, the theory can be consistently
extended to curved spaces \cite{Yano1973} and thus allows for the
inclusion of gravity
\cite{Schuller:2002rr,Schuller:2002fn}.

The first proposals to introduce a maximal acceleration into otherwise
relativistic kinematics were made and pursued by Caianiello and collaborators
\cite{Caianiello, Caia2, Caia3}, using a suitable metric on the tangent bundle of
spacetime. Pseudo-complex spacetime establishes, additionally, a
product structure on the tangent bundle, and hence circumvents the
incompatibility of a maximal acceleration and the strong principle of
equivalence \cite{Schuller:2002rr,Schuller:2002fn}. 

In proposals like \cite{Nest1, Nest2}, the
maximal acceleration is enforced by the particular dynamics of the
investigated models. This is analogous to the r\^ole of the speed of
light in pre-relativistic Maxwell electrodynamics. Before 1905,
the boost symmetry of Maxwell theory was regarded as a dynamical
symmetry beyond the kinematical $SO(3)$ symmetry of the Maxwell
Lagrangian. Using this modern parlance, special relativity can be
regarded as the kinematisation $SO(3) \longrightarrow SO(1,3)$ of the dynamical symmetries of Maxwell
theory that are due to the existence of a universal speed appearing in
the dynamics. Analogously, Born-Infeld kinematics \cite{Schuller:2002rr}
absorbs the dynamical symmetries of Born-Infeld electrodynamics that are
associated with the existence of a maximal electric field strength, into
the spacetime geometry, giving rise to the enlargement of the
transformation group $SO(1,3) \longrightarrow SO_\pc(1,3)$. We review
the technical details of this kinematisation briefly in section \ref{sec_pcM}. 

In an earlier attempt \cite{Nesterenko:1998jt} to study the effects of
a maximal acceleration $\mathfrak{a}$ in quantum field theory, Nesterenko, Feoli, Lambiase and Scarpetta
quantise a sub-maximally accelerated classical point particle in
Minkowski spacetime,
\begin{equation}\label{L_NFLS}
  L_{\textrm{\small NFLS}} = \frac{m}{\mathfrak{a}} \sqrt{\mathfrak{a}^2+{\ddot x}^2}\sqrt{\dot x^2},
\end{equation}
but only consider its standard Poincar\'e symmetry. Such an 
\textit{ad hoc} introduction of the maximal acceleration, through
modified dynamics, inevitably leads to the
higher order derivatives in (\ref{L_NFLS}), so that
the transition to the Hamiltonian formulation must be performed
using the Ostrogradski formalism.
The authors construct 'field equations' of a corresponding second quantised
theory by imposing the classical Noether conservation laws as operator
equations 
on a spacetime function. The Green's function of the arising field equation
is of sixth order, which  leads the authors to the conjecture that a
maximal acceleration principle might possibly regularise quantum field theory.
It is recognised, however, that for the rigorous establishment of such a claim, one needs a field theory based on maximal acceleration \textit{kinematics} from
the outset. In particular, fields must belong to well-defined
irreducible representations of an appropriate symmetry group.

In this paper, we take the $SO_\pc$-symmetry inspired from Born-Infeld theory
seriously as the kinematical symmetry of fundamental physics, and thus show the equivalence of
\begin{itemize}
  \item[(i)] quantum field theory on pseudo-complex
spacetime,
  \item[(ii)] Pauli-Villars regularised quantum field
theory on real spacetime, and
  \item[(iii)] a finite upper bound on
admissible particle accelerations.
\end{itemize}
More precisely, a quantum field theory based on the
pseudo-complexified Poincar\'e group, gives rise to Pauli-Villars
regularised propagators, with a cutoff determined
by the maximal acceleration parameter $\mathfrak{a}$. One understands
this result directly from the representation theory of the pseudo-complex
Poincar\'e group: Each standard particle is found to have as its
'pseudo-partner' a Weyl ghost of equal spin, but generically different
mass. Taking the maximal acceleration parameter to infinity (or,
equivalently, the minimal length to zero)
\textit{after} the calculation of field theoretical amplitudes
completes the Pauli-Villars regularisation prescription, removing the
unitarity-violating ghosts. The necessity to take this limit 
illustrates the fact that quantum field theory is intrinsically a
theory based on the concept of point particles.

% This result establishes
% rigorously the conjecture of Nesterenko, Feoli, Lambiase and
% Scarpetta on the regularising property of the maximal acceleration
% principle, even though via a somewhat different mechanism.\\

% The emerging formalism essentially consists in the substitution of
% all real numbers in standard quantum field theory by pseudo-complex
% numbers. Thus, the theory derives from a clear physical postulate, is
% technically simple, and provides non-trivial results mainly due to
% the existe% nce of zero-divisors in $\pc$. 

The organisation of the paper is as follows. We start by reviewing the
pseudo-complex techniques needed for an understanding of Born-Infeld
kinematics in flat spacetime. In section \ref{sec_reptheory}, we find
the irreducible representations of the
pseudo-complex Poincar\'e group, providing a proper definition of
quantum mechanical particles with sub-maximal
acceleration. The existence of degenerate representations requires the
exclusion of certain fields from the particle spectrum,
which can be achieved by an adaptation of the action principle for
pseudo-complex quantum theories, as explained in section \ref{sec_AP}. 
Dynamics for
the free scalar field representation are explicitly constructed in section
\ref{sec_pcFT}. This leads to a 
pseudo-complex propagator, whose projection to real spacetime is shown
be Pauli-Villars regularised. 
Conversely, in section \ref{sec_phasespace}, we demonstrate that the
field phase space of a Pauli-Villars regularised theory
carries an integrable 
pseudo-complex structure, giving rise to a pseudo-complex field theory.
Section \ref{sec_gaugetheory} deals with the extension of the explicit
constructions to spinor and non-abelian vector fields. In section
\ref{sec_conclusion}, we summarise and conclude.

\section{Pseudo-Complexified Minkowski Space\label{sec_pcM}}
We briefly review the pseudo-complex formulation of Born-Infeld
kinematics, as developed in \cite{Schuller:2002fn, Schuller:2002rr}.
The commutative unit ring of pseudo-complex numbers  over the field
$F \in \{\reals, \mathbb{C}\}$,
\begin{equation}
  \pc_F \equiv \left\{ Q_1 + I Q_2 \,\, | \,\, Q_1, Q_2 \in F \right\},
\end{equation}
where $I \not\in F$, $I^2=+1$, possesses zero divisors
\begin{equation}\label{zerodiv}
  \pc_\pm^0 \equiv \left\{ \lambda(1\pm I)\,\, |\,\, \lambda \in F
  \right\},
\end{equation}
for which no multiplicative inverses exist in $\pc_F$.
For notational convenience, we use the shorthand $\pc \equiv \pc_\reals$. 
The zero divisors will play a crucial r\^ole later on, and it is 
often useful to decompose a pseudo-complex number $Q
= Q_1 + I Q_2 \in \pc_F$ into its zero divisor components 
\begin{equation}
  Q_\pm \equiv Q_1 \pm Q_2 \quad \in F,
\end{equation}
via the multiplicatively acting projectors
\begin{equation}
  \sigma_\pm \equiv \frac{1}{2}(1 \pm I) \in \pc^0_\pm,
\end{equation}
so that
\begin{equation}\label{zd_comp}
  Q = \sigma_+ Q_+ + \sigma_- Q_-.
\end{equation}

A function $f: \pc_F \longrightarrow \pc_F$ is called
\textsl{pseudo-complex differentiable}, if it is, understood as a map
$\tilde f: F^2 \longrightarrow F^2$, $F$-differentiable and satisfies
the pseudo-Cauchy-Riemann equations
%\begin{subequations}
\begin{eqnarray}
  \partial_1 \tilde f_1 &=& \partial_2 \tilde f_2,\label{PCR1}\\
  \partial_2 \tilde f_1 &=& \partial_1 \tilde f_2.\label{PCR2}
\end{eqnarray}
%\end{subequations}
These allow to re-identify $D\tilde f: F^2 \longrightarrow F^{2\times
2}$ with $Df: \pc_F \longrightarrow \pc_F$, where the pseudo-complex
differential operator $D$ can hence be written
\begin{equation}
  D = \frac{1}{2}(\partial_1 + I \partial_2).
\end{equation}

The pseudo-complexification of a finite-dimensional real vector space $M\equiv
\reals^{1+n}$,
\begin{equation}
  M_\pc \equiv \left\{ X \equiv x + I \mathfrak{a}^{-1} u | x,u \in M \right\} = \pc^{1+n},
\end{equation}
where $\mathfrak{a}$ will assume the r\^ole of a fundamental finite
upper bound on accelerations later on,
is a (free) module over $\pc$. Equipping $M_\pc$ with a metric
$\eta$ of signature $(1,n)$ leads to a metric module $\pc^{1,n}$.
A basis $\{e_{(\mu)}\}$ of $M_\pc$ where the metric takes the diagonal
form
\begin{equation}
  \eta(e_{(\mu)},e_{(\nu)}) = \eta_{\mu\nu} \equiv
  \textrm{diag}(1, -1, \dots, -1),
\end{equation}
is called a uniform frame. In such, the symmetry group of
$(M_\pc,\eta)$ is generated by the pseudo-complexified Lorentz algebra
\begin{equation}
  so_\pc(1,n) \equiv \left\{ \omega_{\mu\nu} M^{\mu\nu} | \omega_{\mu\nu} \in \pc, M^{\mu\nu} \in
  so_\reals(1,n)\right\}.
\end{equation}
Clearly, $so_\reals(1,n) \subset so_\pc(1,n)$ is a subalgebra, and
hence pseudo-complex Lorentz-invariant theories do not break
Lorentz invariance.\\
A curve $X: \reals \longrightarrow \pc^{1,n}$ is called an orbit, if
there exists a uniform frame where
\begin{equation}\label{orbitdef}
  X^\mu = x^\mu + I \mathfrak{a}^{-1} \frac{dx^\mu}{d\tau},
\end{equation}
with $d\tau^2 \equiv dx^\mu dx_\mu$ being the real Minkowskian line
element. Such uniform frames are called inertial frames for the orbit $X$.\\
For an orbit $X= x + I \mathfrak{a}^{-1} u$ in an arbitrary uniform frame,
the relation $u = dx/d\tau$ does not generally hold, but is seen 
\cite{Schuller:2002fn} to be
weakened to the orthogonality 
\begin{equation}\label{uaortho}
  \eta(dx,du) = 0.
\end{equation}
Therefore, for an orbit $X$, the $SO_\pc(1,n)$-invariant line element
$d\omega$, defined by
\begin{equation}\label{domega}
  d\omega^2 \equiv dX^\mu\, dX_\mu,
\end{equation}
is always real-valued. This allows the following definition:
An orbit $X$ is called \textsl{sub-maximally accelerated}, if
\begin{equation}\label{submaxacc}
  \eta(dX,dX) > 0.
\end{equation}
For the real spacetime projection $x: \reals \longrightarrow
\reals^{1,n}$ of a sub-maximally accelerated orbit $X = x + I \mathfrak{a}^{-1}
u$, one finds an upper bound on the scalar acceleration,
\begin{equation}\label{accscale}
  - \frac{d^2x^\mu}{d\tau^2} \frac{d^2x_\mu}{d\tau^2} < \mathfrak{a}^2,
\end{equation} 
so that $x$ is a spacetime trajectory of Minkowski curvature less than
$\mathfrak{a}$, justifying the above terminology.

Note that (\ref{accscale}) introduces a Lorentz-scalar
acceleration scale $\mathfrak{a}$, (or, equivalently, a length scale
$\mathfrak{a}^{-1}$) into the theory, which is 
manifestly compatible with an undeformed action of the
Lorentz group on spacetime. It is argued in \cite{Fitz1, Fitz2} that the
introduction of a universal length scale into relativity enforces a
deformation of the Lorentz boosts, due to the Lorentz-Fitzgerald
contraction. The latter, however, only affects directed lengths,
i.e. spatial three-vectors. A (scalar) length scale is a weaker
concept, compatible with standard Lorentz symmetry, as is exemplified by
the Lorentz-invariant condition (\ref{accscale}). In particular, the
present proposal is very different from the type of frameworks
first proposed in
\cite{Amelino-Camelia}.

Using the $SO_\pc(1,n)$-invariant line element (\ref{domega}),
one can combine (\ref{uaortho}) and (\ref{submaxacc}) into the single
scalar condition 
\begin{equation}
  \eta(\frac{dX}{d\omega}, \frac{dX}{d\omega}) = 1,
\end{equation}
for a sub-maximally accelerated orbit $X$. Thus $SO_\pc(1,n)$ manifestly
preserves the orthogonality (\ref{uaortho}) and the maximal
acceleration scale. 

The transformations contained in $SO_\pc(1,n)$ have a clear physical
interpretation as the standard boosts and rotations, and further
transformations to uniformly accelerated and rotating frames. To see
this in detail, consider an observer in Minkowski spacetime, given by
a curve $e_a(\tau)$ in the frame bundle, providing a local orthonormal
basis with $\eta(e_a,e_b) = \eta_{ab}$ at each point of the observer's
worldline $x(\tau)$. Arranging for comoving frames in the standard way, i.e. $e_0 \equiv
\frac{dx}{d\tau}$, where $d\tau$ is the natural parameter of the curve
$x$, we obtain the Frenet-Serret formula \cite{MTW}
\begin{equation}
   \frac{d}{d\tau} e_a = {\theta_a}^b e_b,
\end{equation}     
with the antisymmetric Frenet-Serret tensor $\theta_{ab} = -
\theta_{ba}$, whose $\theta_{0\alpha}$ components encode the
translational three-acceleration $a_\alpha$ of the observer, whereas
the $\theta_{\alpha\beta}$ components describe the angular velocity of
the spatial frame in the $\alpha\beta$--plane, with respect to a
Fermi-Walker transported observer. 
Pseudo-complexified Minkowski spacetime $\pc^{1,n}$ possesses
pseudo-complexified tangent spaces $T_x\pc^{1,n} \cong \pc^{1,n}$,
which induces a pseudo-complex frame bundle in turn. Hence, the real
frame $e_a$ is extended to a pseudo-complex frame
\begin{equation}
  E_a = {\gamma_a}^b (\delta_b^c + \frac{I}{\mathfrak{a}}
  {\theta_b}^c) e_c,
\end{equation}  
lifting the orbit definition (\ref{orbitdef}) to the frame
bundle. Here, $\theta_{bc}$ is the Frenet-Serret tensor of the
observer, and ${\gamma_a}^b$ presents a normalisation factor to
  ensure the normalisation of the pseudo-complex frames $E_a$, 
\begin{equation}\label{Enorm}
  \eta(E_a, E_b) = \eta_{ab}.
\end{equation}  
The matrix-valued $\gamma$-factor is, in the pseudo-complex spacetime
picture, absorbed into the definition of the line element
(\ref{domega}), such that, as one can easily check,
\begin{equation}
  E_0 \equiv \frac{dX}{d\omega}.
\end{equation}
The pseudo-complexified Lorentz group parameterizes the gauge degrees
of freedom for the normalization condition (\ref{Enorm}). It contains
the Lorentz group as a subgroup, $SO(1,n) \subset SO_\pc(1,n)$. Indeed, any
pseudo-complex Lorentz transformation $\Lambda \in SO_\pc(1,n)$
uniquely decomposes into a product 
\begin{equation}\label{polar}
  {\Lambda_a}^b = {Q_a}^c {L_c}^b
\end{equation}
of a real Lorentz transformation $L
\in SO(1,n)$, and a pseudo-complex Lorentz transformation $Q$ with purely
pseudo-imaginary parameters $\omega^*_{\mu\nu} = - \omega_{\mu\nu}$,
\begin{equation}
  Q = \exp(\omega_{\mu\nu} M^{\mu\nu}). 
\end{equation}
The real frames (accompanying an inertial observer with $\theta\equiv0$) provide a perfectly good basis for the pseudo-complexified
tangent spaces $T_x\pc^{1,n} = \langle e_a \rangle_\pc$, but the action of
the pseudo-complex Lorentz group will generate a generic
pseudo-complex frame,
\begin{equation}\label{Etrafo}
  E_a = {\Lambda_a}^b e_b.
\end{equation}  
However, due to the polarisation formula (\ref{polar}), $L$ merely
presents a change of inertial frame, and we can therefore assume,
without loss of generality, that $L$ is the identity, and only study
the action of the transformations of type $Q$. Starting with an
inertial observer, i.e. $\theta\equiv 0$ and
${\gamma_a}^b=\delta_a^b$, a boost in spatial $\beta$-direction, with
purely pseudo-imaginary parameter $I\alpha$, according to (\ref{Etrafo}), effects the change to a
frame of uniform acceleration $\mathfrak{a} \tanh{\alpha}$, as one finds
\begin{equation}
  \theta_{0\beta} = -\theta_{\beta 0} = \mathfrak{a} \tanh{\alpha},
\end{equation}
with all other components vanishing. This clearly respects the
maximal acceleration scale, as $-1 < \tanh{\alpha} < 1$.
The corresponding matrix-valued $\gamma$-factor is found to be
\begin{equation}
  \gamma_{00} = -\gamma_{\beta\beta} = \cosh{\alpha}, \qquad
  \gamma_{\alpha\alpha} = -1, \quad \alpha \neq \beta.
\end{equation}
Similarly, a rotation in the $\alpha\beta$--plane with purely
pseudo-imaginary parameter $I\varphi$, acting on the observer, effects
a change to a uniformly rotating frame of angular velocity
$\mathfrak{a} \tan(\varphi)$, as one finds from (\ref{Etrafo}) that
\begin{equation}
  \theta_{\alpha\beta} = -\theta_{\beta\alpha} = \mathfrak{a} \tan{\varphi},
\end{equation}
with the $\gamma$-factor
\begin{equation}
  \gamma_{\alpha\alpha} = \gamma_{\beta\beta} = -\cos{\varphi}, \qquad
  \gamma_{00} = 1,  
\end{equation}
and all other diagonal entries equal to $-1$.
In  particular, the real Lorentz transformations $L \in SO(1,n)$ act
on the velocity, acceleration, and momentum of a particle exactly as
in standard special relativity. This is possible, as detailed above,
as an acceleration or length scale is a weaker concept than a directed
length, and is therefore not subject to the Lorentz-Fitzgerald
contraction, because one can easily formulate Lorentz-covariant bounds, as in
(\ref{accscale}). 

We conclude that the symmetry group $SO_\pc(1,n)$ contains the real Lorentz
transformations as a subgroup, and further the transformations to
uniformly sub-maximally accelerated and rotating frames. Thus the
12 (real) generators $M^{\mu\nu}, I M^{\mu\nu}$ of $SO_\pc(1,n)$ all
possess a clear physical interpretation. 

Classical Lagrangian dynamics with $SO_\pc(1,n)$ symmetry give rise to
sub-maximally accelerated orbits \cite{Schuller:2002fn}, which we
interpret as classical point particles in $n+1$ dimensions, respecting the maximal
acceleration. In quantum theory, however, the particle spectrum is given by all irreducible
representations of the underlying symmetry group. Therefore, we study the
representation theory of the pseudo-complexified Poincar\'e group in
the next section.

\section{Representation theory of the Pseudo-complex Poincar\'e
group\label{sec_reptheory}}
We show that in a quantum theory with pseudo-complex Poincar\'e
invariance, the standard relativistic particle
spectrum is doubled, providing each real particle with a
pseudo-partner of generically different mass, but equal spin.\\

The pseudo-complexified Poincar\'e algebra $\mathcal{P}_\pc$ in $3+1$
dimensions is generated by  
\begin{equation}
  \mathcal{P}_\pc \equiv \left<P^\mu,
  M^{\alpha\beta}\right>_\pc, 
\end{equation}
where, if acting on spacetime functions, $P_\mu \equiv i D_\mu$ are
the generators of translations in $\pc^{1,3}$, and $M^{\alpha\beta} \equiv
X^\alpha D^\beta - X^\beta D^\alpha$ are the generators of $SO_\pc(1,3)$. Decomposition into zero-divisor
components (\ref{zd_comp}) immediately yields two decoupled real Poincar\'e algebras
\begin{eqnarray}
 \mathcal{P}_\pc &=& \sigma_+ \left<P_+^\mu,
 M_+^{\alpha\beta}\right>_\reals \oplus \sigma_- \left<P_-^\mu,
 M_-^{\alpha\beta}\right>_\reals \nonumber\\
 &=& \sigma_+ \mathcal{P}_\reals \oplus \sigma_- \mathcal{P}_\reals \label{PPdecomp},
\end{eqnarray}
where the sum is direct because $\sigma_+ \sigma_- = 0$.
Thus, \textsl{generically}, a
pseudo-complex particle is labelled by two independent real masses 
and two independent spins or helicities, as follows from the
well-known representation theory of the real Poincar\'e group. 

However, we want to realise the representations by pseudo-complex fields $\phi =
\sigma_+ \phi_+ + \sigma_- \phi_-$. Clearly, the subalgebras $\sigma_\pm
\mathcal{P}_\reals$ act independently on the (real) zero-divisor
components $\phi_\pm$. Hence, $\phi_+$ and $\phi_-$ must belong to real
representations of equal spin,  as otherwise the real and
pseudo-imaginary parts of $\phi$, i.e., $\phi_+ \pm \phi_-$, would not
be algebraically defined. We will see this constraint at work more explicitly below.

As $P_\pm^2$ are Casimir operators of the respective real Poincar\'e
algebras $\sigma_\pm \mathcal{P}_\reals$, the
pseudo-complex operator
\begin{equation}
  P^2 = \sigma_+ P_+^2 + \sigma_- P_-^2
\end{equation}
is a Casimir of $\mathcal{P}_\pc$, and its value $M^2 \in \pc$ is
called the \textsl{pseudo-complex mass} of the representation. We
further define the pseudo-complex Pauli-Ljubanski vector
\begin{equation}
  W_\mu = \frac{1}{2} \epsilon_{\mu\gamma\alpha\beta} P^\gamma M^{\alpha\beta},
\end{equation}
whose zero-divisor components 
\begin{equation}
  W_{\mu\pm} = \frac{1}{2} \epsilon_{\mu\gamma\alpha\beta} P_\pm^\gamma
  M_\pm^{\alpha\beta}
\end{equation}
present the real Pauli-Ljubanski vectors of the two real Poincar\'e
algebras $\sigma_\pm \mathcal{P}_\reals$. 

Representations of the pseudo-complex Poincar\'e algebra fall into
three different classes, which we call \textsl{massive},
\textsl{almost massless}, and \textsl{massless}, according to whether
the pseudo-complex mass is no zero-divisor, a zero-divisor, or zero. We now discuss these cases in turn.

\subsection{Massive case ($M^2 \not\in\pc^0$)}
As in this case $M_\pm^2 >0$, the Casimirs of the respective real Poincar\'e algebras
are given by the squared Pauli-Ljubanski vectors $W_\pm^2 = M_\pm^2
J_\pm^2$, with $J_{\pm i} \equiv \frac{1}{2} \epsilon_{ijk} M_\pm^{jk}$. Clearly,
the squared pseudo-complex Pauli-Ljubanski vector $W^2$ is then a
Casimir of $\mathcal{P}_\pc$, and one observes that the pseudo-complex
spin operator
\begin{equation}
  S^2 \equiv \frac{W^2}{M^2}
\end{equation}
can be written
\begin{equation}\label{spinsum}
  S^2 = \sigma_+ J_+^2 + \sigma_- J_-^2,
\end{equation}
using the identity
\begin{equation}\label{sigmaquotient}
\sigma_\pm \frac{R}{Q} = \sigma_\pm \frac{R_\pm}{Q_\pm}
\end{equation}
for $R,Q \in\pc$ and $Q\not\in\pc^0$.\\
For $\phi = \sigma_+ \phi_+ + \sigma_- \phi_-$ to be algebraically
defined, we need that $J_+^2$ and $J_-^2$ both yield the same real
spin. Then, we see from (\ref{spinsum}) that $S^2$, when acting on a pseudo-complex field
representation, is always real, so that for $S^2 = s(s+1)$, the spin
eigenvalues are half-integer,
\begin{equation}
  s \in \frac{1}{2}\mathbb{N}_0.
\end{equation}
A massive pseudo-complex field therefore gives rise to two real
particles of generically different non-zero masses $M_\pm$, but equal
half-integer spins $s$,
\begin{equation}
  \left|M, s\right>_\pc = \left|M_+, s, +\right>_\reals \oplus \left|M_-, s, -\right>_\reals.
\end{equation}
We have included, into the labelling of the real representations,
the zero divisor branch on which the real particles take their
values. This is necessary, because the pseudo-imaginary unit $I$ presents a
Casimir operator, distinguishing the two real representations
$\sigma_\pm \mathcal{P}_\reals$, because
\begin{equation}
  I \sigma_\pm = \pm \sigma_\pm.
\end{equation}
Later, from the explicit construction of a pseudo-complex quantum field
theory, we will see that $\left|+\right>$ indicates a proper real particle, and $\left|-\right>$ a
Weyl ghost. 
 
\subsection{Almost Massless case ($0 \neq M^2 \in \pc^0$)}
Let, without loss of generality, $P_+^2 = M_+^2 > 0$ and $P_-^2 = M_-^2
= 0$. Then $W_+^2 = M_+^2 J_+^2$. From $W_-^2=P_-^2=P_- W_- = 0$, it
follows that $W_- = \lambda P_-$ for some helicity $\lambda \in
\reals$. The helicity operator can be expressed as
\begin{equation}
  \lambda = - \frac{\mathbf{P}_- .\mathbf{W}_-}{\|\mathbf{P}_-\|^2} =
  \frac{\mathbf{P}_-.\mathbf{J}_-}{\|\mathbf{P}_-\|},
\end{equation}
boldface symbols denoting the spatial parts of the respective four-vectors.  
The spin $S_+^2$ and the helicity $\lambda$ act separately on $\phi_+$
and $\phi_-$, respectively. Again, for a pseudo-complex field $\phi$
to be defined, we need that the helicity of $\phi_-$ equals the spin
of $\phi_+$. 

Thus all physically observed massless particles, possessing
\textsl{half-integer}-valued helicities, can be accommodated in the almost
massless pseudo-complex field representation.
In general, an almost massless pseudo-complex particle gives rise to a
real doublet
\begin{equation}
  \left|M_\pm\sigma_\pm, s\right>_\pc = \left|M_\pm, s, \pm\right>_\reals \oplus \left|0, \lambda=s, \mp\right>_\reals.
\end{equation}
Again, $\left|-\right>$ will be seen to indicate a Weyl ghost, and $\left|+\right>$ a proper
particle, according to the eigenvalue $\pm 1$ of $I$.

\subsection{Massless case ($M^2 = 0$)}
Here we are left with two continuous real helicities $\lambda_\pm$ from

\begin{equation}
  W_\pm = \lambda_\pm P_\pm.
\end{equation}
Particles with continuous helicity are not observed in experiment, and we hence
exclude the massless case from the physical particle spectrum.\\

The representation theory of the pseudo-complex Poincar\'e algebra has
shown that the experimentally observed physical particles occur as the massive and almost
massless representations. A pseudo-complex particle gives rise to a
doublet of real particles of equal spins, but generically different
real masses. From the explicit construction of pseudo-complex quantum
field theory, and its spacetime projection, we will find in the
following sections that the $\left|+\right>$ particles are proper real
particles, for which their $\left|-\right>$ pseudo-partners act as
Pauli-Villars regulators.

\section{Trivial Fields\label{sec_AP}}
From the decomposition (\ref{PPdecomp}) of the pseudo-complex
Poincar\'e algebra, it is clear that for a
field representation $\phi$ that takes values only in the zero-divisor
branches ${\pc_{F}}_+^0$ or
${\pc_{F}}_-^0$, the doublet
of real particles collapses to just \textsl{one} real particle. We
must exclude such fields from the particle spectrum, if we do not want
to get back standard quantum field theory on spacetime as a sector of
the pseudo-complex theory. The way to render solutions of a dynamical
theory meaningless, is to devise equations of motion that are
trivially solved by them. Hence, the appropriate formulation of the
action principle for pseudo-complex quantum field theory is
\begin{equation}\label{Paction}
  \delta S \in \pc_F^0,
\end{equation} 
where $S$ is the action of the pseudo-complex theory at hand. 
This adaptation is particularly natural from the algebraic point of view, given
that the zero divisors of a ring play very much the r\^ole of the zero
in a field. However, the avoidance of a breakdown of the particle
doublets into the standard singlets provides the compelling physical
reason for requiring (\ref{Paction}).

Note that for purely $F$-valued field theory, (\ref{Paction}) reduces to the
standard action principle $\delta S = 0$. We will see that for
non-trivial fields, i.e.,  $\phi \not\in {\pc_{F}}_\pm^0$, one can rewrite
(\ref{Paction}) as an equation.

\section{Pseudo-complex Scalar Field\label{sec_pcFT}}
Now having a clear definition of quantum particles with sub-maximal
acceleration at our disposal, we can formulate dynamical equations for
the free field representations. This is achieved in a standard manner
by imposing classical constraints as operator equations on the
appropriate fields.

The massive and almost massless scalar representations of
$\mathcal{P}_\pc$ have Casimirs
%\begin{subequations}
\begin{eqnarray}
  W^2 &=& 0,\label{Wnull}\\
  P^2 &=& M^2,\qquad M\neq 0\label{PisM}.
\end{eqnarray}
%\end{subequations}
To devise a field equation, 
we impose these as constraints on a field $\phi$ on pseudo-complex
spacetime. Condition (\ref{Wnull}) implies that the field is a
function
\begin{equation}
  \phi: \pc^{1,3} \longrightarrow \pc_F.
\end{equation}  
Applying the operator equation (\ref{PisM}) to the Fourier transformation $\tilde
\phi$ of the field $\phi$, gives
\begin{equation}\label{praeeom}
  (P^2 - M^2) \tilde\phi = 0,
\end{equation}
where for $X \equiv X_{(1)} + I X_{(2)}$,
\begin{equation}\label{pcFourierformula}
  \tilde\phi(P) \equiv \int \textrm{d}^4 X_{(1)}\textrm\,{d}^4
  X_{(2)}\, \phi(X) \exp(-iP^\mu X_\mu). 
\end{equation}
Thus, in position space, we find the pseudo-complex Poincar\'e
invariant field equation
\begin{equation}\label{prelimfield}
  (D^2 + M^2) \phi(X) = 0.
\end{equation}
However, following our reasoning in section \ref{sec_AP}, on the exclusion of trivial fields
$\phi \in {\pc_F}^0$ from the spectrum of meaningful dynamical
solutions, we replace (\ref{prelimfield}) by 
\begin{equation}\label{KGmomspace}
  (D^2 + M^2) \phi(X) \in \pc_F^0.
\end{equation}
The Green's function $G(X-Y)$ for the operator $D^2 + M^2$ is defined
by
\begin{equation}
  (D^2 + M^2) G(X-Y) = (2 \pi)^8 I \delta^{(8)}(X-Y),
\end{equation}
so that, in momentum space representation, 
\begin{equation}\label{PKGprop}
  \tilde G(P) = \frac{I}{P^2 - M^2}.
\end{equation}
Observing that $I \sigma_\pm = \pm \sigma_\pm$, and using the identity
(\ref{sigmaquotient}), one gets
\begin{equation}
  \tilde G(P) = \sigma_+ \frac{1}{P_+^2 - M_+^2} - \sigma_- \frac{1}{P_-^2 - M_-^2}.
\end{equation}
Note that $\tilde G$ is generically pseudo-complex valued. 

In order to
compare this result to standard quantum field theory on real
spacetime, we project the field $\phi: \pc^{1,3} \longrightarrow
\pc_F$ to a spacetime field $\varphi + I \pi: \reals^{1,3}
\longrightarrow \pc_F$. Clearly, this must be done for an observer in an inertial
frame, in order to have a well-defined vacuum for the projected field
theory \cite{Unruh}. Such an inertial projection is obviously given by
mapping the pseudo-imaginary part of the momentum $P$ to zero,
\begin{equation}
  P_{(2)} \mapsto 0.\label{projection1}
\end{equation}
The action of this projection on fields can therefore be
 implemented straightforwardly on the Fourier transform
\begin{equation}
  \tilde\phi: \pc^{1,3} \longrightarrow \pc_F,
\end{equation}
such that the projection to a spacetime field is given by
\begin{equation}
  \tilde \varphi (P_1) + I \tilde \pi(P_1) \equiv \tilde \phi(P_1) : \reals^{1,3}
  \longrightarrow \pc_F.\label{projection2}
\end{equation}
This projection clearly breaks the $\mathcal{P}_\pc$-symmetry down to
$\mathcal{P}_\reals$. However, we will see in section
\ref{sec_phasespace} that the pseudo-complex structure \textsl{survives} as the
geometry of the field phase space $(\varphi,\pi)$ of the projected field
\begin{equation}
  \varphi(x) \equiv \int \textrm{d}^4p\, \tilde\varphi(p) \exp(ip^\mu x_\mu). 
\end{equation}
The projection $\tilde\phi \mapsto \tilde\varphi$ is well-defined under real
Poincar\'e transformations, as the diagram
\begin{center}
\setlength{\unitlength}{3158sp}%
\begingroup\makeatletter\ifx\SetFigFont\undefined%
\gdef\SetFigFont#1#2#3#4#5{%
  \reset@font\fontsize{#1}{#2pt}%
  \fontfamily{#3}\fontseries{#4}\fontshape{#5}%
  \selectfont}%
\fi\endgroup%
\begin{picture}(3687,2278)(1351,-1844)
\thinlines
\put(2326, 89){\vector( 1, 0){1725}}
\put(2326,-1711){\vector( 1, 0){1725}}
\put(1808,-106){\vector( 0,-1){1275}}
\put(5026,-61){\vector( 0,-1){1275}}
\put(2626,239){\makebox(0,0)[lb]{\smash{\SetFigFont{10}{12.0}{\rmdefault}{\mddefault}{\updefault}
$\Lambda \in SO_\reals(1,3)$
}}}
\put(2626,-1561){\makebox(0,0)[lb]{\smash{\SetFigFont{10}{12.0}{\rmdefault}{\mddefault}{\updefault}
$\Lambda \in SO_\reals(1,3)$
}}}
\put(4276, 89){\makebox(0,0)[lb]{\smash{\SetFigFont{10}{12.0}{\rmdefault}{\mddefault}{\updefault}
$\tilde\phi(\Lambda^{-1}(p+If))$
}}}
\put(1576,-1711){\makebox(0,0)[lb]{\smash{\SetFigFont{10}{12.0}{\rmdefault}{\mddefault}{\updefault}
$\tilde\varphi(p)$
}}}
\put(1351, 89){\makebox(0,0)[lb]{\smash{\SetFigFont{10}{12.0}{\rmdefault}{\mddefault}{\updefault}
$\tilde\phi(p+If)$
}}}
\put(4576,-1786){\makebox(0,0)[lb]{\smash{\SetFigFont{10}{12.0}{\rmdefault}{\mddefault}{\updefault}
$\tilde\varphi(\Lambda^{-1}p)$
}}}
\end{picture}
\end{center}
commutes.

For non-trivial $\tilde \phi$, relation (\ref{KGmomspace}) can be rewritten as
an equation,
\begin{equation}
  [P_+^2 - M_+^2][P_-^2 - M_-^2] \tilde \phi(P) = 0.
\end{equation}
Application of the inertial frame projection $\tilde \phi \mapsto \tilde
\varphi$, with $P \mapsto P_{(1)}$, yields
\begin{equation}\label{projectedtheory}
  [P_{(1)}^2 - M_+^2][P_{(1)}^2 - M_-^2] \tilde\varphi(P_{(1)}) = 0,
\end{equation}
revealing higher order dynamics for the projected field $\varphi$. The
corresponding real spacetime propagator reads
\begin{equation}
  \tilde g(p) = \frac{1}{[p^2 - M_+^2][p^2 - M_-^2]},
\end{equation} 
writing $p\equiv P_{(1)}$ for short. This can be brought to the form
\begin{equation}\label{PVprop}
  \tilde g_M(p) \equiv (M_+^2-M_-^2) \tilde g(p) = \frac{1}{p^2-M_+^2}
  - \frac{1}{p^2 - M_-^2}.
\end{equation}
In the special relativity limit $\mathfrak{a} \longrightarrow \infty$,
the propagator 
$\tilde g_M(p)$ must reproduce the standard Klein-Gordon propagator
for a scalar real particle of mass $m$. Therefore, we require that
\begin{equation}
  M_+^2 \stackrel{\mathfrak{a}\rightarrow \infty}{\longrightarrow} m^2,
  \qquad M_-^2 \stackrel{\mathfrak{a}\rightarrow \infty}{\longrightarrow} \infty.\end{equation}
In the almost massless case, $M\in\pc^0_-$, say, this correspondence
  principle fixes the pseudo-complex mass, up to field redefinitions, to
\begin{equation}
  M = \sigma_- \mathfrak{a},
\end{equation} 
as $\mathfrak{a}$ is the only massive parameter available. We therefore \textsl{adopt}, in the massive case
$M\not\in\pc^0$, the pseudo-complex mass
\begin{equation}
  M = \sigma_+ m + \sigma_- \mathfrak{a},
\end{equation}
where $m$ is the mass of the real $\phi_+$ particle. With this
motivated choice, the propagator
for the projected spacetime field $\varphi$,
\begin{equation}
  \tilde g_M(p) = \frac{1}{p^2-m^2} - \frac{1}{p^2-\mathfrak{a}^2},
\end{equation}
is seen to be Pauli-Villars regularised, with the cutoff determined by
the maximal acceleration parameter $\mathfrak{a}$. In particular, note 
that the real representation $\left|M_+,0,+\right>$ is a proper particle, 
while $\left|M_-,0,-\right>$ is a Weyl ghost.\\

We conclude that quantum field theory on pseudo-complex spacetime
gives rise, after an inertial projection to real spacetime, to a
Pauli-Villars regularised quantum field theory on real spacetime. This
proves the implication $(i) \Rightarrow (ii)$ stated in the
introduction. The next section will show that the converse also holds. 

\section{Scalar Field Phase Space\label{sec_phasespace}}
It is worthwhile to investigate the symmetries of the projected
spacetime theory (\ref{projectedtheory}). To this end, consider
Pauli-Villars regularised scalar field theory on real spacetime,
\begin{equation}\label{projLag}
  \mathcal{L} = - \frac{1}{(M_+^2 - M_-^2)}\varphi (\square + M_+^2)(\square + M_-^2) \varphi,
\end{equation}
where $M_+ \ll M_-$ are the masses of the particle and the
regulator, respectively.
De Urries and Julve \cite{deUrries:1998bi} developed a Lorentz-covariant version of
the Ostrogradski formalism for higher-derivative scalar field
theories, and proved its equivalence with the standard non-covariant
approach. Defining $\urries{\pm} \equiv (\square + M_\pm^2)$ and $\julve{+}{-} \equiv
M_+^2 - M_-^2$, the Lagrangian (\ref{projLag}) can be cast into the
form
\begin{eqnarray}
  \mathcal{L} &=& - \frac{1}{{\julve{+}{-}}} \varphi \urries{+}\urries{-} \varphi\nonumber\\
   &\cong& - \frac{1}{{\julve{+}{-}}} \urries{+} \varphi \urries{+}
   \varphi + \varphi \urries{+} \varphi,
\end{eqnarray}
observing that $\urries{-} = \urries{+} - \julve{+}{-}$, and
discarding surface terms. Hence, it
suffices to consider derivatives of the form $\urries{+}\varphi$. 
Defining the canonical momentum density
\begin{equation}
  \pi \equiv \frac{\partial \mathcal{L}}{\partial \urries{+}\varphi}, 
\end{equation}
and solving for $\urries{+}\varphi$ in terms of $\varphi$ and $\pi$, 
\begin{equation}
  \urries{+}\varphi = - \frac{\julve{+}{-}}{2}(\pi - \varphi),
\end{equation}
one obtains the positive definite Hamiltonian density
\begin{eqnarray}\label{Hamdensity}
  \mathcal{H}_1 &\equiv& \pi \urries{+} \varphi - \mathcal{L}(\varphi,
  \urries{+}\varphi(\varphi, \pi))\nonumber\\
  &=& - \frac{\julve{+}{-}}{4} (\pi - \varphi)^2.
\end{eqnarray}
It is shown in \cite{deUrries:1998bi} that the evolution in field
phase space $(\varphi,\pi)$ is then governed by the Hamiltonian equations
%\begin{subequations}
\begin{eqnarray}
  \urries{+} \varphi &=& \frac{\partial \mathcal{H}_1}{\partial \pi}\label{Ham1},\\
  \urries{+} \pi &=& \frac{\partial \mathcal{H}_1}{\partial \varphi}\label{Ham2},
\end{eqnarray}
%\end{subequations}
exhibiting manifestly the \textsl{almost pseudo-complex} structure of
the field phase space of a fourth-order Lagrangian field theory.   
In case the above pair of equations can be combined into one single
pseudo-complex equation, we speak of a \textsl{pseudo-complex}
structure. 

We now identify the necessary and sufficient condition for
an almost pseudo-complex phase space structure (\ref{Ham1}-\ref{Ham2}) to be pseudo-complex. Assume that there exists a real-valued function
$\mathcal{H}_2(\varphi, \pi)$, such that
\begin{equation}\label{fullHext}
  \mathcal{H} \equiv \mathcal{H}_1 + I \mathcal{H}_2
\end{equation} 
satisfies the pseudo-Cauchy-Riemann equations
(\ref{PCR1}-\ref{PCR2}). In this case, we call $\mathcal{H}$ a
pseudo-complex extension of $\mathcal{H}_1$. 
Combining the field and its canonical momentum into one pseudo-complex
valued field on real spacetime,
\begin{eqnarray}
  \phi: \reals^{1,3} &\longrightarrow& \pc,\\
  \phi(x) &\equiv& \varphi(x) + I \pi(x),
\end{eqnarray}
one can write (\ref{Ham1}-\ref{Ham2}) as 
\begin{equation}\label{pcHam}
 \urries{+} \phi = I \frac{D\mathcal{H}}{D\phi},
\end{equation}
if, and only if, $\mathcal{H}$ is a pseudo-complex extension of $\mathcal{H}_1$.
This shows that the field phase space of any fourth order Lagrangian (scalar)
theory possesses a pseudo-complex structure, if, and only if, the
corresponding Hamiltonian has a pseudo-complex extension.

For the particular Hamiltonian (\ref{Hamdensity}), describing a Pauli-Villars
regularised field, such extensions exist and are unique up to an
arbitrary pseudo-complex constant $C$,
\begin{equation}\label{PVHam}
  \mathcal{H} = (1-I) \mathcal{H}_1 + C =
  -\frac{\julve{+}{-}}{2}\sigma_- \phi^2 + C,
\end{equation} 
as can be seen directly from the integration of the pseudo-Cauchy-Riemann
equations for (\ref{fullHext}).

The dynamics of a field theory with pseudo-complex phase space
structure can be captured within the single equation (\ref{pcHam}), 
involving only one field degree of freedom. Therefore, this equation can be
obtained from a Lagrangian,
\begin{equation}
  \mathcal{L} = \frac{1}{2} \phi (\square + M_+^2 ) \phi - I \mathcal{H}(\phi).
\end{equation}
For the special case of a Pauli-Villars regularised spacetime theory,
the potential (\ref{PVHam}) is a mass term, which can be absorbed into
the free Lagrangian,
\begin{eqnarray}\label{pconreal}
  \mathcal{L} = \frac{1}{2}\phi(x)(\square + M^2)\phi(x),
\end{eqnarray}
with a pseudo-complex mass
\begin{equation}
  M = \sigma_+ M_+ + \sigma_- M_-.
\end{equation}
Note that (\ref{pconreal}) describes a pseudo-complex valued field
$\phi$ defined on real, rather than pseudo-complex spacetime. However,
in an inertial frame, this is equivalent to the fully pseudo-complex
Poincar\'e invariant dynamics
\begin{equation}\label{manifestpc}
  \mathcal{L} = \frac{1}{2} \phi(X)(D^2 + M^2)\phi(X)
\end{equation}
on pseudo-complex spacetime. This is most easily seen starting from the Fourier
transform of (\ref{pcHam}),
\begin{equation}\label{almostpc}
  - (p^2 - M_+^2) \tilde\phi(p) = \julve{+}{-} \sigma_- \tilde\phi(p),
\end{equation}
where we have used (\ref{PVHam}). In an inertial frame, the
pseudo-complex extension $p \mapsto P = p + I f$ does not change this
equation, because $f=0$, so that
\begin{equation}
  (P^2 - M^2) \tilde\phi(P) = 0. 
\end{equation}
This is recognised as the Fourier transform (\ref{pcFourierformula}) of the equation of motion derived from the manifestly pseudo-complex
Poincar\'e invariant Lagrangian (\ref{manifestpc}).\\

We conclude that a Pauli-Villars regularised scalar theory on real spacetime
gives rise to a scalar field theory on pseudo-complex spacetime, due
to the integrability of the almost pseudo-complex structure.
This proves the implication $(ii) \Rightarrow (i)$ stated in the
introduction. Together with the results from section \ref{sec_pcFT},
we have thus explicitly shown the equivalence of quantum field theory on
pseudo-complex spacetime and Pauli-Villars regulated quantum field
theory on real spacetime, in the case of a scalar
field. The constructions can be extended to higher
tensor and spinor fields, whose pseudo-complexification we will discuss
in the next section.

\section{Spinor and Vector Fields\label{sec_gaugetheory}}
It is straightforward to apply the pseudo-complexification
procedure to spinor or higher tensor fields. The pseudo-complex Dirac
Lagrangian for an $SO_\pc(1,3)$-spinor $\psi$ reads
\begin{equation}
  \bar \psi(i \gamma^\mu D_\mu - M) \psi,
\end{equation}
with pseudo-complex mass $M\neq 0$, and standard Dirac gamma
matrices. An almost massless, abelian $SO_\pc(1,3)$-vector field
$A^\mu$ is governed by the pseudo-complexified Proca Lagrangian
\begin{equation}
  -\frac{1}{4} F^{\mu\nu}F_{\mu\nu} + \frac{1}{2} M^2 \sigma_- A^\mu A_\mu,
\end{equation} 
where $F_{\mu\nu} \equiv D_\mu A_\nu - D_\nu A_\mu$, and we assume,
without loss of generality, $M \in \pc^0_-$. 

Pauli-Villars regularisation of a vector field in standard quantum
field theory on real spacetime requires the introduction of a non-zero
regulating mass, which breaks gauge invariance. 
We now demonstrate that, in contrast, gauge invariance is fully
preserved in an \textsl{almost massless} pseudo-complex non-abelian gauge theory,
and only broken by the projection (\ref{projection2}) to spacetime.

Consider the pseudo-complexified Dirac Lagrangian
\begin{equation}\label{PDirac}
  \mathcal{L} = \bar\Psi(i \gamma^\mu D_\mu - M) \Psi
\end{equation}
for an $N$-multiplet of spinor fields
\begin{equation}
  \Psi: \pc^{1,3} \longrightarrow \pc_\mathbb{C},
\end{equation}
with pseudo-complex mass $M\neq 0$. Let $\Psi$ belong to an
irreducible representation of a simple compact Lie group $G$, with
generators $\mathbf{t}_a$ satisfying the algebra
\begin{equation}
  [\mathbf{t}^a,\mathbf{t}^b] = i f^{abc} \mathbf{t}^c.
\end{equation}
The theory (\ref{PDirac}) possesses the global gauge symmetry
\begin{equation}\label{globalgauge}
  \Psi \mapsto \exp(i \mathbf{\alpha}^a \mathbf{t}^a) \Psi.
\end{equation}
Now we promote the $\mathbf{\alpha}^a$ to fields $\mathbf{\alpha}^a: \pc^{1,3} \longrightarrow \pc$,
and require (\ref{globalgauge}) to be a local symmetry. Define the gauge covariant derivative
\begin{equation}
  \nabla_\mu \equiv D_\mu - i g A^a_\mu \mathbf{t}^a,
\end{equation} 
where the $A^a: \pc^{1,3} \longrightarrow \pc^{1,3}$ are
taken to be almost massive
vector fields with pseudo-complex mass $N \in \pc^0_-$. The free field
dynamics for the multiplet $A$ is correspondingly given by the Proca-Lagrangian
\begin{equation}\label{freeA}
  \mathcal{L}_A = - \frac{1}{4} F^{a\,\mu\nu}
  F^a_{\mu\nu} + \frac{1}{2}
  N^2 \sigma_- A^{a\,\mu} A^a_\mu,
\end{equation} 
where $F^a_{\mu\nu} \equiv D_\mu A^a_\nu - D_\nu A^a_\mu + g
f^{abc} A^b_\mu A^c_\nu $.
For the covariant derivative to commute with the gauge transformation,
\begin{equation}
  \nabla_\mu(\Psi) \mapsto \exp(i \mathbf{\alpha}^a \mathbf{t}^a) \nabla_\mu \Psi, 
\end{equation} 
we must require that, for infinitesimal $\alpha^a$, 
\begin{equation}\label{Atrafo}
  A^a_\mu \mathbf{t}^a \mapsto A^a_\mu \mathbf{t}^a + \frac{1}{g}
  \left(D_\mu \mathbf{\alpha}^a\right) \mathbf{t}^a + i \alpha^a A^b_\mu
  f^{abc} \mathbf{t}^c 
\end{equation}
under a local gauge transformation. The full Lagrangian
\begin{equation}
  \mathcal{L} = \bar\Psi(i \gamma^\mu \nabla_\mu - M) \Psi - \mathcal{L}_A 
\end{equation}
is then seen to be gauge invariant if, and only if, we constrain the gauge
parameters to zero-divisor values $\pc^0_+$,
\begin{equation}
  \alpha^a: \pc^{1,3} \longrightarrow \pc^0_+ \cong \reals,
\end{equation} 
so that the change of $A^a_\mu \mathbf{t}^a$ in (\ref{Atrafo}) is $\pc^0_+$-valued, and therefore the mass term in (\ref{freeA}) is gauge
invariant. Hence, none of the gauge symmetry present in standard
non-abelian gauge theory is lost. 

This shows that it is merely the spacetime formulation of standard
quantum field theory that causes the conflict between the gauge
principle and Pauli-Villars regularisation.
It is an open question as to what extent one can exploit
this symmetry, e.g., obtain Ward identities, in pseudo-complex quantum
field theory. The investigation of such questions requires a careful
analysis of non-trivial interaction effects due to the existence of
zero-divisors. It should be interesting to address these questions in
future research. 

\section{Conclusion\label{sec_conclusion}}
On Dirichlet branes in ten-dimensional superstring theory,
electrically charged particles can experience an acceleration of at most the order of the inverse string length, as the
electrodynamics are governed by the Born-Infeld action, with its
well-known maximal field strength. Brane world
scenarios, assigning to Dirchlet three-branes the r\^ole of the
observed four-dimensional universe, therefore suggest that there is an
upper limit to accelerations in such models.   
 
As shown in earlier work \cite{Schuller:2002rr, Schuller:2002fn}, the kinematics of a
relativistic particle with sub-maximal acceleration can be encoded in the
pseudo-complexified Poincar\'e group. Its representation
theory reveals that a pseudo-complex quantum particle gives rise to a doublet
of real particles, with different real masses but equal spin, if
described by a pseudo-complex field theory. Exactly one of these
particles is identified as a Weyl ghost that acts as a
Pauli-Villars regulator of the other, proper particle. Hence, in
pseudo-complex quantum field theory, particles always carry their
regulators around.

 Removal of the new fundamental acceleration (or,
length) scale by means of taking the appropriate limit \textsl{after}
the quantization, then renders the resulting theory unitary. Thus, it
seems that a classical theory with invariant acceleration (length)
scale naturally gives rise to a better-behaved quantum field theory
than standard relativity. This insight also sheds new light on
  the necessity to remove the Pauli-Villars regulator at the end of
  calculations: this corresponds to taking the string length to zero,
  thus returning to a standard quantum field theory of point particles.  

The abstract results from the representation theory are confirmed by
the explicit construction of the theory for a scalar field, which is
governed by a pseudo-complexified Klein-Gordon equation. 
The pseudo-complex denominator of the
corresponding scalar propagator generates a double infinity of poles
due to the existence of zero divisors. An analysis of its
projection to a spacetime field confirms that this is equivalent to a
Pauli-Villars regularisation.  Remarkably, the pseudo-complex structure of the full
theory, which is broken by the spacetime projection, resurfaces as the
geometrical structure of the phase space for the regularised spacetime
field.  
We find that the Pauli-Villars regularisation of a real
spacetime theory induces a pseudo-complex field theory, and vice versa. 
This equivalence between maximal acceleration kinematics, 
pseudo-complex quantum field theory, and  
Pauli-Villars regularisation,
rigorously establishes a conjecture \cite{Nesterenko:1998jt} by
Nesterenko, Feoli, Lambiase and Scarpetta.

The extension to spinor and vector fields is straightforward. 
Pseudo-complex gauge theory features the standard gauge symmetry, 
although it projects to a Pauli-Villars regularised theory on spacetime.
An in-depth analysis of interacting pseudo-complex quantum field
theory remains to be done in future work, where particular attention should
be paid to non-trivial effects due to the existence of zero divisors
in $\pc$.

%\appendix
%\section{}

\ack
  We thank Gary Gibbons and Paul Townsend for helpful discussions. 
  FPS is funded by the Studienstiftung des deutschen Volkes, an Isaac Newton
  scholarship of the Cambridge European Trust, and EPSRC. MNRW acknowledges
  financial support by the Gates Cambridge Trust. TWG is funded by the
  Studienstiftung des deutschen Volkes.  

%\bibliography{pvr_bib}

\section*{References}
\begin{thebibliography}{00}
\bibitem{Born:1934gh}
 Born M and Infeld L 1934
 \textit{Proc. Roy. Soc. Lond.} \textbf{A144} 425

\bibitem{Abu87}
 Abuelsaood \etal 1987 \textit{Nucl. Phys.} \textbf{B280} 599; Callan \etal 1988 \textit{Nucl. Phys.} \textbf{B308} 221

\bibitem{Randall}
 Randall L and Sundrum R 1999 \textit{Phys. Rev. Lett} \textbf{83} 4690

\bibitem{Schuller:2002rr}
 Schuller F P 2002 \textit{Phys. Lett.} \textbf{B540} 119

\bibitem{Schuller:2002fn}
 Schuller F P 2002 \textit{Annals
Phys.} \textbf{299} 174

\bibitem{Thomas}
 Thomas L H 1927 {\textit Nature} \textbf{117} 1

\bibitem{Newman} Newman D, Ford G W, Rich A and Sweetman E 1978 \textit{Phys. Rev. Lett.}
\textbf{40} 1355

\bibitem{Amelino-Camelia}
 Amelino-Camelia G 2002 \textit{Int. J. Mod. Phys.} {\bf D11} 35

\bibitem{Kowalski-Glikman:2002ft}
Kowalski-Glikman J 2002
\textit{Phys. Lett.} \textbf{B547} 291
%%CITATION = HEP-TH 0207279;%%

\bibitem{Yano1973} Yano K and Ishihara S 1973 \textit{Tangent and Cotangent
Bundles} (New York: Marcel Dekker)

\bibitem{Caianiello} E.R. Caianiello 1981 \textit{Lett. N. Cimento} \textbf{32}, 65 

\bibitem{Caia2} E.R. Caianiello, S.de Filippo, G. Marmo, G. Vilasi
  1982 \textit{Lett. N. Cimento} \textbf{34},
112 

\bibitem{Caia3} E.R. Caianiello, M. Gasperini, G. Scarpetta 1991 \textit{Class. Q. Grav.} \textbf{8}, 659 

\bibitem{Nest1} V. V. Nesterenko, A. Feoli, and G. Scarpetta 1996 \textit{Class.Quant.Grav.} \textbf{13} 1201-1212

\bibitem{Nest2} V. V. Nesterenko 1993 \textit{J.Math.Phys.} \textbf{34} 5589-5595

\bibitem{Fitz1}
Amelino-Camelia G 2001 \textsl{Phys. Lett.} \textbf{B510} 255-263 

\bibitem{Fitz2}
Kowalski-Glikman J and Nowak S 2002 \textsl{Phys. Lett.} \textbf{B539} 126-132

\bibitem{Nesterenko:1998jt}
 Nesterenko V V, Feoli A, Lambiase G and Scarpetta G 1999
 \textit{Phys. Rev.} \textbf{D60} 065001

\bibitem{MTW} C. W. Misner, K. S. Thorne and J. A. Wheeler, Gravitation, W. H. Freeman 1973. 

\bibitem{Unruh} 
Unruh W G 1968 \textit{Phys. Rev.}  
\textbf{D14} 
870

\bibitem{deUrries:1998bi}
de Urries F J and Julve J 1998
 \textit{J. Phys.} \textbf{A31} 6949
\end{thebibliography}

%%CITATION = HEP-TH 9802115;%%

\end{document}